\documentclass[11pt,draftclsnofoot,letterpaper,onecolumn]{IEEEtran}
\usepackage{amsmath}
\usepackage{amssymb}
\usepackage{epsfig}
\usepackage{verbatim}

\DeclareSymbolFont{calsymbols}{OMS}{cmsy}{m}{n} \DeclareSymbolFontAlphabet{\mathcal}{calsymbols}

\def\E{\mathop{\cal E}\nolimits}             
\def\I{\mathop{\cal I}\nolimits}            
\newcommand{\Linf}{\mathcal{L}_{\infty}}
\newcommand{\Sinf}{\mathcal{S}_{\infty}}

\begin{document}

\newtheorem{theorem}{Theorem} \newtheorem{lemma}{Lemma}
\newtheorem{definition}{Definition} \newtheorem{corollary}{Corollary}
\newtheorem{proposition}{Proposition}

\title{High SNR Analysis for MIMO Broadcast Channels:
Dirty Paper Coding vs.~Linear Precoding}

\author{\authorblockN{Juyul Lee and Nihar Jindal}\\
\authorblockA{Department of Electrical and Computer Engineering\\
University of Minnesota\\
Minneapolis, MN 55455, USA\\
E-mail: \{juyul,nihar\}@umn.edu}}

\maketitle

\begin{abstract}
We study the MIMO broadcast channel and compare the achievable throughput for the optimal strategy of dirty
paper coding to that achieved with sub-optimal and lower complexity linear precoding (e.g., zero-forcing and
block diagonalization) transmission.  Both strategies utilize all available spatial dimensions and therefore
have the same multiplexing gain, but an absolute difference in terms of throughput does exist.  The sum rate
difference between the two strategies is analytically computed at asymptotically high SNR, and it is seen
that this asymptotic statistic provides an accurate characterization at even moderate SNR levels.
Furthermore, the difference is not affected by asymmetric channel behavior when each user a has different
average SNR. Weighted sum rate maximization is also considered, and a similar quantification of the
throughput difference between the two strategies is performed. In the process, it is shown that allocating
user powers in direct proportion to user weights asymptotically maximizes weighted sum rate. For multiple
antenna users, uniform power allocation across the receive antennas is applied after distributing power
proportional to the user weight.
\end{abstract}


\section{Introduction}

The multiple antenna broadcast channel (BC) has recently been the subject of tremendous interest, primarily
due to the realization that such a channel can provide MIMO spatial multiplexing benefits without requiring
multiple antenna elements at the mobile devices \cite{Caire_IT03}. Indeed, it is now well known that dirty
paper coding (DPC) achieves the capacity region of the multiple antenna BC \cite{Weingarten_IT06}. However,
implementation of DPC requires significant additional complexity at both transmitter and receiver, and the
problem of finding practical dirty paper codes that approach the capacity limit is still unsolved.

On the other hand, linear precoding is a low complexity but sub-optimal transmission technique  (with
complexity roughly equivalent to point-to-point MIMO) that is able to transmit the same number of data
streams as a DPC-based system.  Linear precoding therefore achieves the same multiplexing gain (which
characterizes the slope of the capacity vs.~SNR) curve) as DPC, but incurs an absolute rate/power offset
relative to DPC.  The contribution of this work is the quantification of this rate/power offset.

In this work, we apply the high SNR affine approximation \cite{Shamai_IT01} to the sum rate capacity (DPC)
and to the linear precoding sum rate. Both approximations have the same slope (i.e., multiplexing gain), but
by characterizing the difference in the additive terms the rate/power offset between the two strategies is
determined.  By averaging the per-channel realization rate offset over the iid Rayleigh fading distribution
we are able to derive very simple expressions for the average rate offset as a function of only the number of
transmit and receive antennas and users for systems in which the aggregate number of receive antennas is no
larger than the number of transmit antennas.

Note that previous work has analyzed the \textit{ratio} between the sum rate capacity and the linear
precoding sum rate \cite{Jindal_IT05_DPCvsTDMA}\cite{Shen_ISIT06}.  In this work we alternatively study the
\textit{absolute difference} between these quantities, which appears to be a more meaningful metric precisely
because both strategies provide the same multiplexing gain.

In addition to sum rate, we also study weighted sum rate maximization (using DPC and linear precoding) and
provide simple expressions for the rate offsets in this scenario. One of the most interesting results is that
weighted sum rate (for either DPC or linear precoding) is maximized at asymptotically high SNR  by
\textit{allocating power directly proportional to user weights.} A similar result was recently observed in
\cite{Lozano_Tulino_Verdu_ISSSTA} in the context of parallel single-user channels (e.g., for OFDMA systems).
Because the linear precoding strategies we study result in parallel channels, the result of
\cite{Lozano_Tulino_Verdu_ISSSTA} shows that it is asymptotically optimal to allocate power in direct
proportion to user weights whenever linear precoding is used.  By showing that weighted sum rate maximization
when DPC is employed can also be simplified to power allocation over parallel channels, we are able to show
that the same strategy is also asymptotically optimal for DPC. To illustrate the utility of this simple yet
asymptotically optimal power allocation policy, we apply it to a system employing queue-based scheduling (at
finite SNR's) and see that it performs extremely close to the true optimal weighted sum rate maximization.

This paper is organized as follows: Section II presents the system model and Section III introduces the high
SNR capacity approximation from \cite{Shamai_IT01}. Section IV describes dirty paper coding and linear
precoding and derives simple expressions for their sum rates at high SNR, and in Section V the relative
rate/power offsets between DPC and linear precoding are computed. Section VI extends the analysis to weighted
sum rate maximization and considers a queue-based system with the weighted sum rate solution. We conclude in
Section VII.

\section{System Model}
We consider a $K$-user Gaussian MIMO BC in which the transmitter has $M$ antennas and each receiver has $N$
antennas with $M \geq KN$, i.e., the number of transmit antennas is no smaller than the aggregate number of
receive antennas. The received signal ${\bf y}_k$ of user $k$ is given by
\begin{equation} \label{eq:y_Hx_n}
    {\bf y}_k = {\bf H}_k {\bf x}+{\bf n}_k, \qquad k=1,\cdots, K,
\end{equation}
where ${\bf H}_k(\in {\mathbb{C}}^{N \times M})$ is the channel gain matrix for user $k$, ${\bf x}$ is the
transmit signal vector having a power constraint ${\rm{tr}}(E[{\bf{xx}}^H ]) \le P$, and ${\bf n}_k$
$(k=1,\cdots,K)$ is complex Gaussian noise with unit variance per vector component (i.e., $E[{\bf n}_k^H{\bf
n}_k] = {\bf I}$). We assume that the transmitter has perfect knowledge of all channel matrices and each
receiver has perfect knowledge of its own channel matrix. For the sake of notation, the concatenation of the
channels is denoted by ${\bf H}^H = [{\bf H}_1^H \;{\bf H}_2^H\; \cdots \; {\bf H}_K^H](\in
\mathbb{C}^{M\times KN})$, which can be decomposed into row vectors as ${\bf H}^H = [{\bf h}_{1,1}^H \;{\bf
h}_{1,2}^H\; \cdots \; {\bf h}_{1,N}^H\; {\bf h}_{2,1}^H \;{\bf h}_{2,2}^H\; \cdots \; {\bf
h}_{2,N}^H\;\cdots\; {\bf h}_{K,N}^H]$, where ${\bf h}_{k,n} (\in \mathbb{C}^{1\times M})$ is the $n$th row
of ${\bf H}_k$. We develop rate offset expressions on a per realization basis as well as averaged over the
standard iid Rayleigh fading distribution, where the entries of ${\bf H}$ are iid complex Gaussian with unit
variance.

\emph{Notations}: Boldface letters denote matrix-vector quantities. The operations $\text{tr}(\cdot)$ and
$(\cdot)^H$ represents the trace and the Hermitian transpose of a matrix, respectively. The operations
$|\cdot|$ and $\|\cdot\|$ denote the determinant of a matrix and the Euclidean norm of a vector,
respectively. The operations $\E[\cdot]$ and $\I(\cdot)$ denote the expectation and the mutual information,
respectively.

\section{High SNR Approximation}

This section describes the key analytical tool used in the paper, namely the
affine approximation to capacity at high SNR developed by Shamai and Verd\'u
\cite{Shamai_IT01}.
At high SNR, the channel capacity $C(P)$ is well approximated by an affine function
of SNR ($P$):
\begin{eqnarray}
    C(P) &=& \Sinf \left(\log_2 P -\Linf \right) +o(1)  \nonumber \\
&=& \Sinf \left( \frac{P_{dB}}{3 dB}  - \Linf\right) +o(1),
\label{eq:cap_affine}
\end{eqnarray}
where $\Sinf$ represents the multiplexing gain,
$\Linf$ represents the power offset (in 3 dB units), and
the $o(1)$ term vanishes as $P \rightarrow \infty$.
The multiplexing gain $\Sinf$
and the power offset $\Linf$ are defined as:
\begin{eqnarray}
\Sinf &=& \lim_{P \rightarrow \infty}
\frac {C(P)}{\log_2(P)}, \\
\Linf &=& \lim_{P \rightarrow \infty}
\left( \log_2(P) - \frac{C(P)}{\Sinf} \right).
\end{eqnarray}

This high SNR approximation is studied for point-to-point MIMO channels in
\cite{Lozano_IT05}.  In this context the multiplexing gain $\Sinf$  is well known to equal
the minimum of the number of transmit and receive antennas, and thus is
essentially independent of the fading environment.  However, the rate
offset $\Linf$  does depend on the actual fading
statistics (and possibly on the level of channel state information
available to the transmitter as well), and \cite{Lozano_IT05} provides
exact characterizations of these offset
terms for the most common fading models, such as iid Rayleigh fading,
spatially correlated fading, and Ricean (line-of-sight) fading.
Indeed, one of the key insights of \cite{Lozano_IT05} is the necessity to
consider these rate offset terms, because considering only the multiplexing
 gain can lead to rather erroneous conclusions, e.g., spatial correlation
does not affect MIMO systems at high SNR.

In a similar vein, in this work we utilize the high SNR approximation to quantify the difference between
optimal dirty paper coding and simpler linear precoding in an iid Rayleigh fading environment. The
multiplexing gain is easily seen to be the same for both strategies, but a non-negligible difference exists
between the rate offsets.  By investigating the differential offsets between these two strategies, we are
able to very precisely quantify the throughput degradation that results from using linear precoding rather
than the optimal DPC strategy in spatially white fading\footnote{Although we do not pursue this avenue in the
present publication, it would also be interesting to investigate the DPC-linear precoding offset in other
fading environments, e.g., Ricean and spatially correlated fading.  However, one must be careful with respect
to channel models because some point-to-point MIMO models do not necessarily extend well to the MIMO
broadcast channel.  For example, in point-to-point channels spatial correlation captures the effect of sparse
scattering at the transmitter and/or receiver and is a function of the angle-of-arrival.  In a broadcast
channel, the angle-of-arrival is typically different for every receiver because they generally are not
physically co-located; as a result, using the same correlation matrix for all receivers is not well motivated
in this context.}.

Although the high SNR approximation is exact only at asymptotically high SNR, it is seen to provide very
accurate results for a wide range of SNR values, e.g., on the order of 5 dB and higher. Because multi-user
MIMO systems generally provide a substantial advantage over point-to-point systems (e.g., TDMA-based systems)
at moderate and high SNR's, the approximation is accurate in the range of interest.

\section{Dirty Paper Coding vs.~Linear Precoding}

In this section we compute the affine approximation to the dirty paper coding sum rate and the linear
precoding sum rate using the high SNR approximation.

\subsection{Dirty Paper Coding}

Dirty paper coding (DPC) is a pre-coding technique that allows for pre-cancellation of interference at the
transmitter.  Costa introduced DPC  while considering an AWGN channel with additive interference known
non-causally at the transmitter but not at the receiver \cite{Costa_IT83}.  DPC was applied to the MIMO
broadcast channel, where it can be used to pre-cancel multi-user interference, by Caire and Shamai and was
shown to achieve the sum capacity of the 2-user, $M>1$, $N=1$ MIMO broadcast channel \cite{Caire_IT03}. The
optimality of DPC was later extended to the sum capacity of the MIMO broadcast channel with an arbitrary
number of users and antennas \cite{Vishwanath_IT03}\cite{Viswanath_IT03}\cite{Yu_IT04_SumCapBC}, and more
recently has been extended to the full capacity region \cite{Weingarten_IT06}.

We now describe the transmit signal when DPC is utilized. Let ${\bf
s}_k (\in \mathbb{C}^{N\times 1})$ be the $N$-dimensional vector of
data symbols intended for user $k$ and ${\bf V}_k (\in
\mathbb{C}^{M\times N})$ be its precoding matrix.  Then the transmit
signal vector ${\bf x}$ can (roughly) be represented as
\begin{equation} \label{eq:dirty_paper_sum}
    {\bf x} = {\bf V}_1 {\bf s}_1 \oplus ({\bf V}_2 {\bf s}_2 \oplus   \cdots \oplus ({\bf V}_{K-2} {\bf s}_{K-2} \oplus ({\bf V}_{K-1} {\bf s}_{K-1} \oplus {\bf V}_K {\bf s}_K))\cdots),
\end{equation}
where $\oplus$ represents the non-linear \emph{dirty paper sum}. Here we have assumed, without loss of
generality, that the encoding process is performed in descending numerical order. Dirty-paper decoding at the
$k$-th receiver results in cancellation of ${\bf V}_{k+1} {\bf s}_{k+1}, \ldots, {\bf V}_{K} {\bf s}_{K}$,
and thus the effective received signal at user $k$ is:
\begin{equation} \label{eq:dpc_y}
    \tilde{\bf y}_k = {\bf H}_k {\bf V}_k {\bf s}_k + \sum_{j=1}^{k-1} {\bf H}_k {\bf V}_j {\bf s}_j + {\bf n}_k,
\end{equation}
where the second term is the multi-user interference that is not
cancelled by DPC.  If the ${\bf s}_k$ are chosen Gaussian, user
$k$'s rate is given by:
\begin{eqnarray}
    R_k
    &=& \log_2 \frac{\left|{\bf I}+{\bf H}_k\left(\sum_{j=1}^k {\boldsymbol \Sigma}_j \right) {\bf H}_k^H\right|}{\left|{\bf I}+{\bf H}_k\left(\sum_{j=1}^{k-1} {\boldsymbol \Sigma}_j \right) {\bf
    H}_k^H\right|},
\end{eqnarray}
where ${\boldsymbol \Sigma}_j = {\bf V}_j E[{\bf s}_j {\bf s}_j^H] {\bf V}_j^H$ denotes the transmit
covariance matrix of user $j$. Since DPC is optimal, the sum capacity of the MIMO BC can be expressed as:
\begin{equation} \label{eq-sum_cap_dpc}
    {\cal C}_\text{DPC}({\bf H},P) = \max_{\sum_k \textrm{tr}({\boldsymbol \Sigma}_k)\le P}
    \sum_{k=1}^K \log_2 \frac{\left|{\bf I}+{\bf H}_k\left(\sum_{j=1}^k {\boldsymbol \Sigma}_j \right) {\bf H}_k^H\right|}{\left|{\bf I}+{\bf H}_k\left(\sum_{j=1}^{k-1} {\boldsymbol \Sigma}_j \right) {\bf
    H}_k^H\right|},
\end{equation}
where the maximization is performed over the transmit covariance
matrices ${\boldsymbol \Sigma}_1, {\boldsymbol \Sigma}_2, \cdots,
{\boldsymbol \Sigma}_K$.

The duality of the MIMO broadcast channel and the MIMO multiple
access channel (MAC) allows the sum capacity to alternatively be
written as \cite{Vishwanath_IT03}:
\begin{equation} \label{eq:sum_cap}
    {\cal C}_\text{DPC}({\bf H},P) = \max_{\sum_k \textrm{tr}({\bf Q}_k)\leq P} \log_2
    \left|{\bf I}+\sum_{k=1}^K   {\bf H}_k^H{\bf Q}_k {\bf H}_k
    \right|,
\end{equation}
where ${\bf Q}_k$ represent the $N \times N$ transmit covariance
matrices in the dual MAC.

No closed-form solution to either (\ref{eq-sum_cap_dpc}) or to
(\ref{eq:sum_cap}) (which is a convex problem) is known to exist,
but it has been shown that ${\cal C}_\text{DPC}({\bf H},P)$
converges (absolutely) to the capacity of the point-to-point MIMO
channel with transfer matrix ${\bf H}$ whenever $M \geq KN$:
\begin{theorem}[Theorem 3 in \cite{Caire_IT03}] \label{thm:sum_rate_dpc}
    When $M\ge KN$ and ${\bf H}$ has full row rank,
    \begin{equation}
        \lim_{P\to\infty} \left[{\cal C}_\text{DPC}({\bf H},P)
-\log_2\left|{\bf I}+ \frac{P}{KN} {\bf H}^H {\bf H} \right| \right]=0.
    \end{equation}
\end{theorem}

We are now able to make a few important observations regarding the
optimal covariance matrices at high SNR.  Since
\begin{equation} \label{eq:sum_cap_mod}
    \log_2 \left|{\bf I}+ \frac{P}{KN} \sum_{k=1}^K   {\bf H}_k^H  {\bf H}_k
    \right| = \log_2\left|{\bf I}+ \frac{P}{KN} {\bf H}^H {\bf H}
    \right|,
\end{equation}
choosing each of the dual MAC covariance matrices as ${\bf Q}_k = \frac{P}{KN} {\bf I}$ in \eqref{eq:sum_cap}
achieves sum capacity at asymptotically high SNR.  Thus, uniform power allocation across the $KN$ antennas in
the dual MAC is asymptotically optimal. It is also possible to determine the optimal form of the downlink
covariance matrices ${\boldsymbol \Sigma}_1, \ldots, {\boldsymbol \Sigma}_K$, or equivalently of the downlink
precoding matrices ${\bf V}_1, \ldots, {\bf V}_K$.  When $N=1$, Theorem 3 of \cite{Caire_IT03} shows that a
technique referred to as \emph{zero-forcing DPC} asymptotically achieves sum capacity. Zero-forcing DPC,
which is implemented via the QR-decomposition of the concatenated channel matrix ${\bf H}$ in
\cite{Caire_IT03}, implies that the precoding matrices ${\bf V}_1, \ldots, {\bf V}_K$ are chosen to
completely eliminate multi-user interference, and thus to satisfy ${\bf H}_k {\bf V}_j = 0$ for all $j < k$.
Because DPC eliminates some of the multi-user interference terms, ${\bf V}_1$ has no constraint on it, ${\bf
V}_2$ must be orthogonal to ${\bf H}_1$, ${\bf V}_3$ must be orthogonal to ${\bf H}_1$ and ${\bf H}_2$, and
so forth. If multi-user interference is eliminated, then the system decouples into $K$ parallel channels, and
simply using equal power allocation across all of the channels is asymptotically optimal due to the well
known fact that waterfilling over parallel channels provides no advantage at high SNR.

As a result of Theorem \ref{thm:sum_rate_dpc}, an affine approximation for the sum rate can be found as:
\begin{equation}
    {\cal C}_\text{DPC}({\bf H},P) \cong KN\log_2 P -KN\log_2 KN+
\log_2\left|{\bf H}{\bf H}^H \right|,
    \label{eq:sum_rate_dpc_appr}
\end{equation}
where $\cong$ refers to equivalence in the limit (i.e., the difference between both sides converges to zero
as $P\to\infty$). Since the MIMO broadcast and the $M \times KN$ point-to-point MIMO channel are equivalent
at high SNR, the high SNR results developed in \cite{Lozano_IT05} directly apply to the sum capacity of the
MIMO broadcast channel.  It is important to be careful regarding the ordering of the equivalent
point-to-point MIMO channel: due to the assumption that $M \geq KN$, the MIMO broadcast is equivalent to the
$M \times KN$ MIMO channel \textit{with CSI at the transmitter}, which is equivalent to the $KN \times M$
MIMO channel \textit{with or without CSI at the transmitter}.  When $M > KN$, the level of CSI at the
transmitter affects the rate offset of the $M \times KN$ point-to-point MIMO channel. Finally, notice that
the high SNR sum rate capacity only depends on the product of $K$ and $N$ and not on their specific values;
this is not the case for linear precoding.

\subsection{Linear Precoding}

Linear precoding is a low-complexity, albeit sub-optimal, alternative to DPC.
When linear precoding is used, the transmit signal vector ${\bf x}$ is a
linear function of the symbols
 ${\bf s}_k (\in \mathbb{C}^{N\times 1}),\;\;k=1,\cdots,K$:
\begin{equation}
    {\bf x}=\sum_{k=1}^K {\bf V}_k {\bf s}_k,
\end{equation}
where ${\bf V}_k (\in\mathbb{C}^{M\times N})$ is the precoding matrix for user $k$. This expression
illustrates linear precoding's complexity advantage: if DPC is used, the transmit signal is formed by
performing dirty-paper sums, which are complex non-linear operations, whereas linear precoding requires only
standard linear operations. The resulting received signal for user $k$ is given by
\begin{equation}
    {\bf y}_k = {\bf H}_k {\bf V}_k {\bf s}_k + \sum_{j\ne k}{\bf H}_k {\bf V}_j {\bf s}_j + {\bf n}_k,
    \label{eq:y_precoding}
\end{equation}
where the second term in \eqref{eq:y_precoding} represents the
multi-user interference. If single-user detection and Gaussian
signalling are used, the achievable rate of user $k$ is:
\begin{equation}
    R_k = \I ({\bf s}_k ; {\bf y}_k)
    = \log_2 \frac{\left|{\bf I}+{\bf H}_k\left(\sum_{j =1}^{K} {\boldsymbol \Sigma}_j \right) {\bf H}_k^H\right|}
    {\left|{\bf I}+{\bf H}_k\left(\sum_{j \ne k} {\boldsymbol \Sigma}_j \right) {\bf
    H}_k^H\right|}.
\end{equation}
Since DPC is not used, each user is subject to multi-user
interference from every other user's signal.  As a result, the
precoding matrices must satisfy very stringent conditions in order
to eliminate multi-user interference.  Note that eliminating
multi-user interference is desirable at high SNR in order to prevent
interference-limited behavior.

In this paper we consider two linear precoding schemes that
eliminate multi-user interference when $M\ge KN$: zero-forcing (ZF)
and block diagonalization (BD). The precoding matrices $\{{\bf
V}_j\}_{j=1}^K$ for BD are chosen such that for all $j (\ne k) \in
[1, K]$,
\begin{equation}
    {\bf H}_k {\bf V}_j = {\bf O},
\end{equation}
while those for ZF are chosen so that
\begin{eqnarray}
    &{\bf h}_{k,n} {\bf v}_{j,l} = 0, &\quad \forall j (\ne k) \in [1, K], \;\; \forall n, l\in [1, N],\\
    &{\bf h}_{k,n} {\bf v}_{k,l} = 0, &\quad \forall l (\ne n) \in [1, N],
\end{eqnarray}
where ${\bf v}_{j,l}$ denotes the $l$th column vector of ${\bf V}_j$. Consequently, performing ZF in a system
with $K$ users with $N(>1)$ antennas is equivalent to performing ZF in a channel with $KN$ single antenna
receivers. Note that ${\bf H}$ having full row rank is sufficient to ensure ZF and BD precoding matrices
exist. In iid Rayleigh fading ${\bf H}$ has full row rank with probability one.

\subsubsection{Zero-forcing}
When ZF is employed, there is no multi-user or inter-antenna interference. Then the received signal at the
$n$th antenna of user $k$ is given by
\begin{equation}
    y_{k,n} = {\bf h}_{k,n} {\bf v}_{k,n} {s}_{k,n} + n_{k,n}, \qquad n=1, \cdots, N,
\end{equation}
where $s_{k,n}$ and $n_{k,n}$ denote $n$th component of ${\bf s}_k$ and ${\bf n}_k$, respectively. Thus, ZF
converts the system into $KN$ parallel channels with effective channel $g_{k,n} ={\bf h}_{k,n} {\bf
v}_{k,n}$. Sum rate is maximized by optimizing power allocation across these parallel channels:
\begin{equation} \label{eq:c_sum_zf}
    {\cal C}_\text{ZF}({\bf H},P)= \max_{\sum_k\sum_n P_{k,n}\le P}
    \sum_{k=1}^{K} \sum_{n=1}^N \log_2 \left(1+P_{k,n}|g_{k,n}|^2 \right).
\end{equation}
Since the optimum power allocation policy converges to uniform power at asymptotically high SNR
\cite{Jindal_ISIT05}, we have:
\begin{equation}
    {\cal C}_\text{ZF}({\bf H},P) \cong KN\log_2 P -KN\log_2 KN+ \log_2\prod_{k=1}^K\prod_{n=1}^N |g_{k,n} |^2.
    \label{eq:sum_rate_zf_appr}
\end{equation}
This approximation is identical to that for DPC in (\ref{eq:sum_rate_dpc_appr}) except for the final constant
term.

\subsubsection{Block Diagonalization}
When BD is employed, there is no multi-user interference because the precoding matrix for BD is chosen to be
${\bf H}_k {\bf V}_j = {\bf O}$ for $k\ne j$. Then the received signal for user $k$ is given by
\begin{equation}
    {\bf y}_k = {\bf H}_k {\bf V}_k {\bf s}_k + {\bf n}_k.
\end{equation}
Thus, BD converts the system into $K$ parallel MIMO channels with effective channel matrices  ${\bf G}_k={\bf
H}_k{\bf V}_k$, $k=1,\cdots,K$. The BD sum rate is given by \cite{Choi_WIRELESS04}\cite{Spencer_SP04}
\begin{equation} \label{eq:sum_bd}
    {\cal C}_\text{BD} ({\bf H},P) = \max\limits_{{\bf Q}_k : \sum\limits_{k=1}^K \text{tr}\{{\bf Q}_k\} \le P}\;\;\;\sum_{k=1}^K \log_2 \left|{\bf I}+ {\bf G}_k{\bf Q}_k{\bf
        G}_k^H \right|,
\end{equation}
and the optimal rate is achieved asymptotically by uniform power allocation at high SNR since the channel can
be decomposed into parallel channels. Hence, the sum rate is asymptotically given by
\begin{equation}
    {\cal C}_\text{BD} ({\bf H},P) \cong KN \log_2 P - KN\log_2 KN + \log_2 \prod_{k=1}^K |{\bf G}_k^H{\bf
    G}_k|. \label{eq:sum_rate_bd_appr}
\end{equation}

\subsection{Equivalent MIMO Interpretation}
\label{sec-equiv_mimo}

 Due to the properties of iid Rayleigh fading,
systems employing either zero-forcing or block diagonalization are equivalent to parallel point-to-point MIMO
channels, as shown in \cite{Choi_WIRELESS04}.  When ZF is used, the precoding vector for each receive antenna
(i.e., each row of the concatenated channel matrix ${\bf H}$) must be chosen orthogonal to the other $KN-1$
rows of ${\bf H}$.  Due to the isotropic nature of iid Rayleigh fading, this orthogonality constraint
consumes $KN-1$ degrees of freedom at the transmitter, and reduces the channel from the $1 \times M$ vector
${\bf h}_{k,n}$ to a $1 \times (M - KN + 1)$ Gaussian vector.  As a result, the effective channel norm
$|g_{k,n}|^2$ of each parallel channel is chi-squared with $2(M-KN+1)$ degrees of freedom (denoted
$\chi^2_{2(M-KN+1)}$). Therefore, a ZF-based system with uniform power loading is exactly equivalent (in
terms of ergodic throughput) to $KN$ parallel $(M-KN+1) \times 1$ MIMO channels (with CSIT).

When BD is used, the orthogonality constraint consumes $(K-1)N$ degrees of freedom.  This reduces the channel
matrix ${\bf H}_k$, which is originally $N \times M$, to a $N \times (M - (K-1)N)$ complex Gaussian matrix.
As a result, the $N \times N$ matrix ${\bf G}_k^H {\bf G}_k$ is Wishart with $M-(K-1)N$ degrees of freedom,
and therefore a BD-based system is equivalent to $K$ parallel $(M- (K-1)N) \times N$ parallel MIMO channels
(with CSIT).

Finally, when DPC is used, the MIMO broadcast channel is equivalent to the $M \times KN$ point-to-point MIMO
channel, where $M \geq KN$ and CSIT is again assumed. Note that a MIMO channel of this dimension can be
interpreted as a series of parallel channels as well: in this case, the $M \times KN$ channel is equivalent
to $M \times 1$, $M - 1 \times 1$, \ldots, $M - KN + 1 \times 1$ channels in parallel \cite{Foschini_Gans}.

For all three cases, the MIMO equivalence is exact when uniform
power loading is used with ZF ($P_{k,n} = \frac{P}{KN}$ for all $k,n$ in
(\ref{eq:c_sum_zf})), BD
(${\bf Q}_k = \frac{P}{KN} {\bf I}$ for all $k$ in (\ref{eq:sum_bd})),
and DPC ( ${\bf Q}_k = \frac{P}{KN} {\bf I}$ for all $k$ in (\ref{eq:sum_cap})).
If optimal power allocation is performed, for either ZF, BD, or DPC,
the MIMO broadcast systems can achieve a larger ergodic throughput
than the MIMO equivalent at finite SNR.
However, because waterfilling provides a vanishing benefit as SNR is increased,
this advantage disappears at asymptotically high SNR.

\begin{figure}
    \centering
    \includegraphics[width=1.0\textwidth]{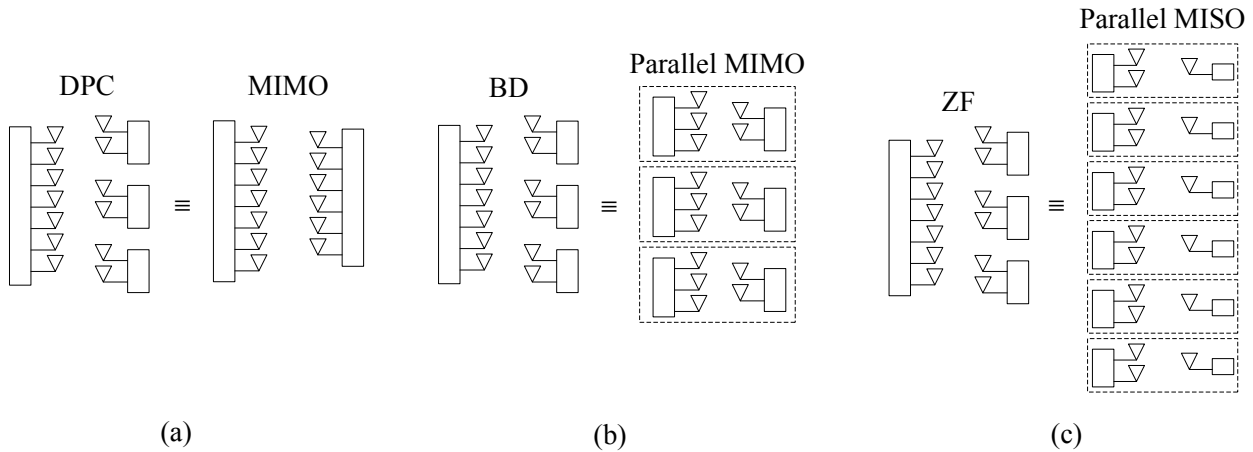}
    \caption{The broadcast channel with $M=7$, $N=2$, and $K=3$ can be interpreted in terms of its sum rate as (a) $7\times 6$ point-to-point MIMO channel when
    DPC is employed, (b) 3 parallel $3\times 2$ MIMO channels when BD is employed, and (c) 6 parallel $2\times 1$ MISO channels when ZF
    is employed.}
    \label{fig:summary_sum_rates}
\end{figure}

The equivalent MIMO channels are summarized in Table
\ref{tbl:summary_sum_rates} and illustrated in
Fig.~\ref{fig:summary_sum_rates} for $M=7$, $N=2$, $K=3$.  In this
case ZF is equvalent to 6 parallel $2 \times 1$ channels, BD is
equivalent to 3 parallel $3 \times 2$ channels, and DPC is
equivalent to a $7 \times 6$ channel.  The absolute difference in
throughput at asymptotically high SNR is indeed due to the diference
in the degrees of freedom in the available parallel channels, as
made precise in the following section.

\begin{table}
    \centering
    \caption{Sum rates at high SNR and their equivalent MIMO interpretation}
    \label{tbl:summary_sum_rates}
    \begin{tabular}{|c|c|c|}
        \hline
                & ${\cal C}({\bf H},P)$ &  MIMO Interpretation \\
        \hline \hline
            DPC & $\log\left|\frac{P}{KN}{\bf H}^H{\bf H}\right|$ &  one $M\times KN$ \\
        \hline
            BD & $\sum_{k=1}^K \log \left| \frac{P}{KN} {\bf G}_k^H {\bf G}_k\right|$ &  $K$ parallel $(M-(K-1)N)\times N$ \\
        \hline
            ZF & $\sum_{k=1}^{K} \sum_{n=1}^N \log\left(\frac{P}{KN} |g_{k,n}|^2\right)$ &  $KN$ parallel $(M-KN+1)\times 1$ \\
        \hline
    \end{tabular}
\end{table}

Our analysis is limited to channels in which $M \geq KN$.  If $M < KN$, i.e., there are strictly less
transmit antennas than aggregate receive antennas, then no MIMO equivalent channel exists for either DPC or
linear precoding.  The sum capacity (DPC) is smaller than the capacity of the $M \times KN$ (forward)
cooperative channel (in which CSIT is not required at high SNR), but is larger than the capacity of the
reverse $KN \times M$ cooperative channel without CSIT.  Zero forcing and block diagonalization are clearly
only feasible when the number of data streams is no greater than $M$.  Thus, if there are more than $M$
receive antennas, some form of selection (of users and possibly of the number of data streams per receiver)
must be performed. As a result of these complications, it does not appear that the high SNR framework will
yield closed-form solutions for either DPC or linear precoding when $M < KN$.

\section{Sum Rate Analysis}

This section quantifies the sum rate degradation incurred by linear precoding relative to DPC.  In terms of
the high SNR approximation, this rate offset is essentially equal to the difference between the $\Linf$ terms
for DPC and linear precoding.

\subsection{DPC vs.~ZF}

We define the rate loss as the asymptotic (in SNR) difference between the sum rate capacity and the zero
forcing sum rate:
\begin{equation}
    \beta_\text{DPC-ZF}({\bf H}) \triangleq \lim_{P\to\infty} \left[{\cal C}_\text{DPC}({\bf
    H},P) - {\cal C}_\text{ZF}({\bf H},P)\right].
\end{equation}
Since each of the capacity curves has a slope of $\frac{KN}{3}$ in
units of bps/Hz/dB, this rate offset
(i.e., the vertical offset between capacity vs. SNR curves) can be
immediately translated into a power offset (i.e., a horizontal
offset): $\Delta_\text{DPC-ZF}({\bf
H})=\frac{3}{KN}\beta_\text{DPC-ZF}({\bf H})$ dB. Because
$\Delta_\text{DPC-ZF}$ is in dB units, we clearly have
$\Delta_\text{DPC-ZF}({\bf H})=3({\cal
L}_\infty^\text{ZF}({\bf H})-{\cal L}_\infty^\text{DPC}({\bf H}))$, which implies
\begin{eqnarray}
\Linf^\text{ZF} &=& \Linf^\text{DPC} + \frac{1}{3} \Delta_\text{DPC-ZF} \\
&=& \Linf^\text{DPC} + \frac{1}{KN} \beta_\text{DPC-ZF} \label{eq-linf_zf}
\end{eqnarray}

From the affine approximation to DPC and ZF sum rate found in \eqref{eq:sum_rate_dpc_appr} and
\eqref{eq:sum_rate_zf_appr}, the rate loss incurred by ZF is:
\begin{equation}
    \beta_\text{DPC-ZF}({\bf H})= \log_2 \frac{|{\bf H}^H{\bf H}|}{\prod_{k=1}^{K}\prod_{n=1}^N |g_{k,n}|^2}.
\end{equation}
While the above metric is the rate loss per realization, we are more interested in the average rate offset
across the fading distribution:
\begin{eqnarray} \label{eq-dpc-zf_offset_defn}
\bar{\beta}_\text{DPC-ZF} = \E_{\bf H} \left[\beta_\text{DPC-ZF}({\bf H}) \right],
\end{eqnarray}
which allows a comparison of ergodic (over the fading distribution) throughput. Likewise, the average power
offset is denoted as $\bar{\Delta}_\text{DPC-ZF}$ and can be immediately calculated in the same fashion.
Under iid Rayleigh fading, the matrix ${\bf H}{\bf H}^H$ is Wishart with $2M$ degrees of freedom while
$|g_{k,n}|^2$ are identically $\chi^2_{2(M-KN+1)}$, as explained in Section \ref{sec-equiv_mimo}.

The key to computing the average offset is the following closed form
expression for the expectation of the log determinant of a Wishart matrix:
\begin{lemma}[Theorem 2.11 of \cite{Tulino_Book04}] \label{lem-wishart}
If $m \times m$ matrix ${\bf H}{\bf H}^H$ is complex Wishart distributed with $n$ ($\geq m$) degrees of
freedom (d.o.f), then:
\begin{equation}
    \E\Big[\log_e \left|{\bf H}{\bf H}^H \right|\Big] = \sum_{l=0}^{m-1}\psi(n-l),
\end{equation}
where $\psi(\cdot)$ is Euler's digamma function,  which satisifes
\begin{eqnarray}
\psi(m) = \psi(1) + \sum_{l=1}^{m-1} \frac{1}{l}
\end{eqnarray}
for positive integers $m$ and $\psi(1)\approx -0.577215$.
\end{lemma}
This result can be directly applied to chi-squared random variables by noting that a $1 \times 1$ complex
Wishart matrix with $n$ degrees of freedom is $\chi^2_{2n}$:
\begin{equation}
    \E[\log_e \chi^2_{2n}] = \psi(n).
\end{equation}

Using Lemma \ref{lem-wishart} we can compute the average rate offset
in closed form:
\begin{theorem} \label{thm:exp_beta_zf}
    The expected loss in Rayleigh fading due to zero-forcing is
    given by
    \begin{equation}
        \boxed{\bar{\beta}_\text{DPC-ZF}(M,KN) =\log_2 e \sum_{j=1}^{KN-1} \frac{j}{M-j}\quad \text{(bps/Hz)}. }\label{eq:E_beta_zf}
    \end{equation}
\end{theorem}
\vspace{5pt}
\begin{proof}
Since ${\bf H}{\bf H}^H$ is $KN \times KN$ Wishart with $M$ d.o.f, and $|g_{k,n}|^2$ is $\chi^2_{2(M-KN+1)}$,
Lemma 1 applied to (\ref{eq-dpc-zf_offset_defn}) gives:
\begin{eqnarray}
\bar{\beta}_\text{DPC-ZF} &=&\E\left[\log_e |{\bf H}^H{\bf H}|
\right] - KN \cdot \E\left[\log_e |g_{1,1}|^2\right] \\
&=&  \log_2 e \left[ \left(\sum_{l=0}^{KN-1}\psi(M-l)\right)-KN \psi(M-KN+1) \right] \label{eq-dpc_zf_proof1}
\end{eqnarray}
By expanding the digamma function and performing the algebraic manipulations shown in Appendix
\ref{sec:pf_thm:exp_beta_zf}, the form \eqref{eq:E_beta_zf} can be reached.
\end{proof}

Using this result we easily get an expression for the rate offset $\Linf^\text{ZF}(M,KN)$ by plugging into
\eqref{eq-linf_zf}
\begin{eqnarray}
\Linf^\text{ZF}(M,KN) &=& \Linf^\text{DPC}(M,KN) +  \frac{1}{KN}
\bar{\beta}_\text{DPC-ZF}(M,KN) \\
&=& \Linf^\text{MIMO}(KN,M) + \frac{\log_2 e}{KN} \sum_{j=1}^{KN-1} \frac{j}{M-j},
\end{eqnarray}
where $\Linf^\text{MIMO}(KN,M)$ is the rate offset of a $KN$ transmit antenna, $M$ receive antenna MIMO
channel in iid Rayleigh fading, which is defined in Proposition 1 of \cite{Lozano_IT05}.

When the total number of receive antennas is equal to $M$ (i.e., $M=KN$),
ZF incurs a rather large loss relative to DPC that can be approximated as:
\begin{equation} \label{eq:beta_zf_M_K_equal}
    \bar{\beta}_\text{DPC-ZF} (M) \approx M\log_2 M \quad \text{(bps/Hz)}
\end{equation}
in the sense that the ratio of both sides converges to one as $M$ grows large
(see Appendix \ref{sec:pf_eq:beta_zf_M_K_equal} for the proof).
 In this scenario, the ZF sum
rate is associated with the capacity of $M$ parallel $1 \times 1$ (SISO)
channels while the DPC sum rate is associated with a $M\times M$
MIMO channel. This corresponds to a power offset of $3\log_2 M$ (dB),
which is very significant when $M$ is large.
Note that the approximation $3\log_2 M$ (dB) overstates the power penalty by 1 to 1.5 dB for reasonable values of
$M(<20)$, but does capture the growth rate. Such a large penalty is not surprising, since the use of
zero-forcing requires inverting the $M\times M$ matrix $\bf H$, which is poorly conditioned with high
probability when $M$ is large.

We can also consider the asymptotic ZF penalty when the number of transmit
antennas is much larger than the number of receive antennas.
If the number of users and transmit antennas are taken to infinity at a
fixed ratio according to $M=\alpha KN$ with $KN\to\infty$ for some $\alpha >1$,
then the power offset between DPC and ZF converges to a constant:
\begin{theorem} \label{thm:asymp_zf_penalty}
    For $M=\alpha KN$ with $\alpha >1$, $KN\to \infty$, the asymptotic power penalty
    for ZF is given by
    \begin{equation} \label{eq:penalty_zf}
        \bar{\Delta}_\text{DPC-ZF}(\alpha) = -3 \left(\log_2 e + \alpha \log_2 \left(1-\frac{1}{\alpha} \right) \right)\qquad\text{(dB)}.
    \end{equation}
\end{theorem}
\begin{proof}
    See Appendix \ref{sec:pf_thm_asymp_zf_penalty}.
\end{proof}
This power offset is incurred due to the fact that the DPC sum rate increases according to a $KN\times \alpha
KN$ MIMO channel capacity while the ZF sum rate increases according to $KN$ parallel $(\alpha-1)KN \times 1$
MISO channels. For example, if $\alpha=2$, or the number of transmit antennas is double the number of
receivers, the zero-forcing penalty is no larger than 1.67 dB, and monotonic convergence to this asymptote is
observed. Thus for large systems, ZF is a viable low-complexity alternative to DPC if the number of transmit
antennas can be made suitably large. A similar conclusion was drawn in \cite{Hochwald_Allerton02} where the
ratio of the rates achievable with ZF relative to the sum capacity is studied. Note that using ZF on the MIMO
downlink channel is identical to using a decorrelating receiver on the multiple antenna uplink channel or in
a randomly spread CDMA system; as a result Theorem \ref{thm:asymp_zf_penalty} is identical to the asymptotic
performance of the decorrelating CDMA receiver given in Eq.~(152) of \cite{Shamai_IT01}.

\begin{figure}
    \centering
    \includegraphics[width=0.6\textwidth]{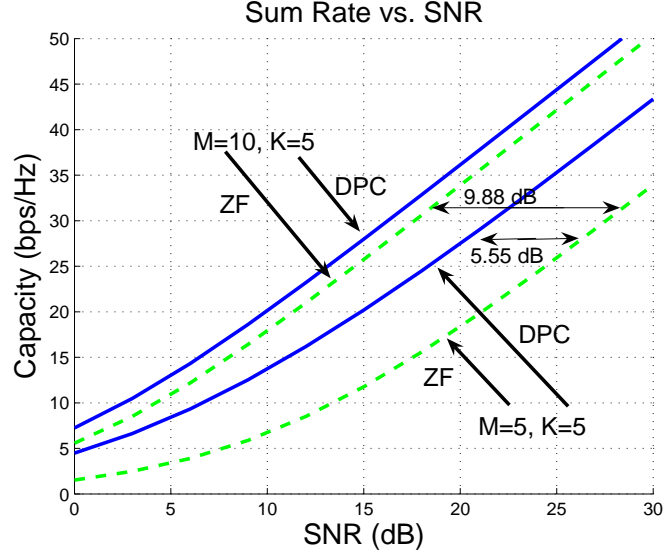}
    \caption{DPC vs.~zero-forcing at high SNR}
    \label{fig:dpc_zf_linf_55_105}
\end{figure}

Figure \ref{fig:dpc_zf_linf_55_105} plots the ZF and DPC throughputs for two five receiver systems. In a
five-transmit-antenna/five-receiver system ($M=K=5, N=1$), Theorem \ref{thm:exp_beta_zf} gives a throughput
penalty of 9.26 bps/Hz, which is equivalent to a power penalty of 5.55 dB (whereas the approximation in
\eqref{eq:beta_zf_M_K_equal} gives 6.97 dB). Although this penalty is exact only in the asymptote, the figure
shows that it gives accurate results throughout the entire SNR range. Throughput curves for a $(M=10, K=5,
N=1)$ system are also shown. The ZF power penalty is only 1.26 dB, which is reasonably close to the
asymptotic penalty of 1.67 dB given by Theorem \ref{thm:asymp_zf_penalty} for $\alpha=2$. Increasing the
number of transmit antennas from 5 to 10 shifts the sum capacity curve by 5.59 dB, but improves the
performance of ZF by 9.88 dB.
 This is because ZF gains the increase in the ${\cal L}_\infty$ term of sum capacity,
along with the significantly decreased ZF penalty due to the increased
number of transmit antennas (5.55 dB to 1.26 dB).
Thus adding transmit antennas has the dual benefit of increasing the
performance of DPC as well as reducing the penalty of
using low-complexity ZF.

\subsection{DPC vs.~BD}

We similarly define the rate loss between DPC and BD as:
\begin{equation}
    \beta_\text{DPC-BD}({\bf H}) \triangleq \lim_{P\to\infty} \left[{\cal C}_{\rm DPC}({\bf H},P)-{\cal C}_{\rm BD}({\bf H},P)\right],
\end{equation}
and denote the expected loss as $\bar{\beta}_\text{DPC-BD}\triangleq \E_{{\bf H}}[\beta_\text{DPC-BD}({\bf
H})]$. Similar to the analysis for ZF, we can calculate the loss terms for a fixed channel and also average
over Rayleigh fading.  In order to compute the average rate loss, we use the fact
that the BD sum rate is asymptotically equal to the capacity of $K$
parallel $N\times (M-(K-1)N)$ iid Rayleigh MIMO channels \cite{Choi_WIRELESS04}.
\begin{theorem} \label{thm:exp_beta_bd}
    The expected loss in Rayleigh fading due to block
    diagonalization is given by
    \begin{equation}
        \boxed{
        \bar{\beta}_\text{DPC-BD} (M, K, N) = (\log_2 e)\left(
        \sum_{k=0}^{K-1}\sum_{n=0}^{N-1}\sum_{i=kN+1}^{(K-1)N}\frac{1}{M-n-i}
        \right) \;\;\; \text{\rm (bps/Hz)}.} \label{eq:E_beta_bd}
    \end{equation}
\end{theorem}
\begin{proof}
    See Appendix \ref{sec:pf_thm:exp_beta_bd}.
\end{proof}
Eq.~\eqref{eq:E_beta_bd} simplifies to \eqref{eq:E_beta_zf} when $N=1$; i.e., zero-forcing is a special case
of block diagonalization. If the number of transmit antennas $M$ is kept fixed but $N$ is increased and $K$
is decreased such that $KN$ is constant, i.e., the number of antennas per receiver is increased but the
aggregate number of receive antennas is kept constant, then the rate offset decreases. In the degenerate case
$M=N$ and $K=1$ the channel becomes a point-to-point MIMO channel and the offset is indeed zero. Using the
same procedure as for ZF, we can easily get an expression for the rate offset $\Linf^\text{BD}(M,K,N)$
(\ref{eq-linf_zf})
\begin{eqnarray}
\Linf^\text{BD}(M,K,N) &=& \Linf^\text{MIMO}(KN,M) +  \frac{1}{KN} \bar{\beta}_\text{DPC-BD}(M,N,K).
\end{eqnarray}

\begin{figure}
    \centering
    \includegraphics[width=0.6\textwidth]{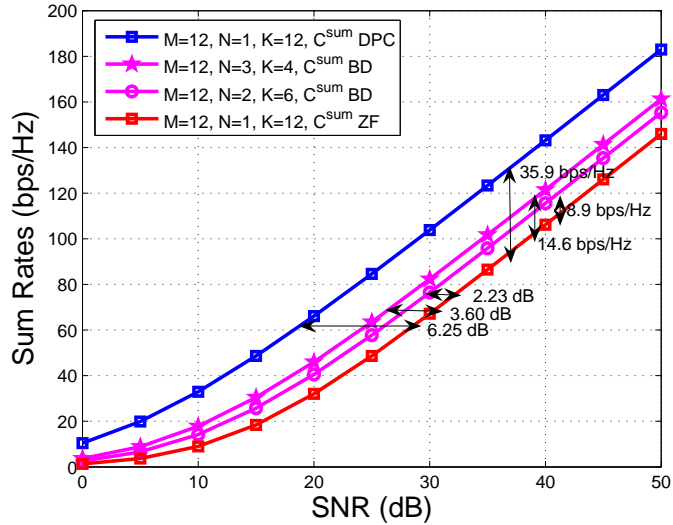}
    \caption{Comparison of \eqref{eq:E_beta_bd} and \eqref{eq:bd_zf_offset_M_KN_equal} with simulated rate losses and power
    offsets}
    \label{fig:beta_bd_zf_offset}
\end{figure}

Although it is difficult to obtain insight directly from
Theorem \ref{thm:exp_beta_bd}, it is much more useful to consider the
offset between BD ($K$ receivers with $N$ antennas each) and
ZF (equivalent to $KN$ receivers with 1 antenna each).
\begin{theorem} \label{thm:diff_beta}
    If $M=\alpha KN$ with $N>1$ and $\alpha > 1$,
the expected throughput gain of BD relative to ZF is:
    \begin{eqnarray*}
        \bar{\beta}_\text{BD-ZF} &\triangleq& \bar{\beta}_\text{DPC-ZF}(M,NK) -
\bar{\beta}_\text{DPC-BD}(M,N,K) \nonumber \\
&=& (\log_2 e)K\sum_{j=1}^{N-1} \frac{(N-j)}{(\alpha-1)KN+j} \quad \text{\rm (bps/Hz)}
\label{eq:bd_zf_diff} \\
&=& \frac{3 \log_2 e}{N} \sum_{j=1}^{N-1} \frac{(N-j)}{(\alpha-1)KN+j} \quad \text{\rm (dB)}.
    \end{eqnarray*}
\end{theorem}
\begin{proof}
    See Appendix \ref{sec:pf_thm:diff_beta}.
\end{proof}
A direct corollary of this is an expression for the expected power offset when $M=KN$:
\begin{equation}
    \boxed{
    \bar{\Delta}_\text{BD-ZF}(N) = \frac{3(\log_2 e)}{N}\sum_{j=1}^{N-1}\frac{N-j}{j}\quad \text{\rm (dB)}.}
    \label{eq:bd_zf_offset_M_KN_equal}
\end{equation}
Note that this expression only depends on the number of receive antennas per receiver and is independent of
the number of users, i.e., of the system size.  For example, consider two system configurations: (i)
$\frac{M}{2}$ receivers each have two antennas, and (ii) $M$ receivers each have one antenna. Equation
\eqref{eq:bd_zf_offset_M_KN_equal} indicates that the power advantage of using BD in the $N=2$ system is
$\bar{\Delta}_\text{BD-ZF}=2.1640$ (dB) relative to performing ZF.  Since this offset is independent of $M$,
it is the same for $M=4$ and $K=4, N=1$ vs.~$K=2, N=2$ systems as well as for $M=6$ and $K=6, N=1$ vs.~$K=3,
N=2$ systems. To illustrate the utility of the asymptotic rate offsets, sum rates are plotted in
Fig.~\ref{fig:beta_bd_zf_offset} for systems with $M=12$ and $N=3$, $K=4$, and $N=2$, $K=6$. Notice that the
asymptotic offsets provide insight at even moderate SNR levels (e.g., 10 dB). When $M=12,\; N=3,\; K=4$,
$\bar{\beta}_\text{BD-ZF}=14.4270$ (bps/Hz) and $\bar{\Delta}_\text{BD-ZF} = 3.6067$ (dB) while the numerical
values are 14.6 (bps/Hz) and 3.65 (dB), respectively.

\subsection{Unequal Average SNR's}

The underlying assumption beforehand is that the strengths of channel gains are the same for all users.
However, there exist near-far effects in a typical wireless broadcast channel scenario which lead to
asymmetric channel gains. In this subsection, we consider the effect of asymmetric channel gains or unequal
average SNR  and reformulate the rate offsets \eqref{eq:E_beta_zf} and \eqref{eq:E_beta_bd}.

We assume that the channel gain of each user can be decomposed into
\begin{equation} \label{eq:H_with_gamma}
    {\bf H}_k = \sqrt{\gamma_k} \tilde{\bf H}_k,\quad k=1,\cdots,K,
\end{equation}
where $\gamma_k$ denotes the average SNR of user $k$. The elements of $\tilde{\bf H}_k$ have Gaussian
distribution with mean zero and unit variance. Notice that the quantities with tilde $\tilde{(\cdot)}$ are
derived under a zero mean unit Gaussian assumption. Then the channel model \eqref{eq:y_Hx_n} is changed to
\begin{equation}
    {\bf y}_k = \sqrt{\gamma_k} \tilde{\bf H}_k{\bf x}_k + {\bf n}_k.
\end{equation}

In the preceding discussion, we have used the fact that the uniform power allocation is asymptotically
optimal for DPC at high SNR. It is important to note that the uniform power allocation is still
asymptotically optimal even when users' SNR are asymmetric. Since ${\bf
H}=(\text{diag}(\sqrt{\gamma_1},\cdots,\sqrt{\gamma_K})\otimes {\bf I}_{N\times N}) \tilde{\bf H}$ where
$\tilde{\bf H}^H = [\tilde{\bf H}_1^H \;\tilde{\bf H}_2^H\; \cdots \; \tilde{\bf H}_K^H]$ and $\otimes$
denotes the Kronecker product, the aggregate channel ${\bf H}$ is full rank with $M\ge KN$. Thus, Theorem
\ref{thm:sum_rate_dpc} holds. When ZF or BD is used, the effective channels are simply multiplied by the
corresponding $\sqrt{\gamma_k}$.

From \eqref{eq:sum_cap}, \eqref{eq:c_sum_zf}, and \eqref{eq:sum_bd} with uniform power allocation, we can
derive the sum rates for asymmetric channel gains as follows
\begin{eqnarray}
    {\cal C}_\text{DPC}({\bf H},P) &\cong& {\cal C}_\text{DPC}(\tilde{\bf H},P) + N \sum_{k=1}^K \log_2 \gamma_k, \label{eq:C_dpc_gamma}\\
    {\cal C}_\text{ZF}({\bf H},P) &\cong& {\cal C}_\text{ZF}(\tilde{\bf H},P) + N \sum_{k=1}^K \log_2 \gamma_k,\label{eq:C_zf_gamma}\\
    {\cal C}_\text{BD}({\bf H},P) &\cong& {\cal C}_\text{BD}(\tilde{\bf H},P) + N \sum_{k=1}^K \log_2 \gamma_k, \label{eq:C_bd_gamma}
\end{eqnarray}
where ${\cal C}_\text{DPC}(\tilde{\bf H},P)$, ${\cal C}_\text{ZF}(\tilde{\bf H},P)$, and ${\cal
C}_\text{BD}(\tilde{\bf H},P)$ are the sum rates under the symmetric channel gain scenario. The derivation of
\eqref{eq:C_dpc_gamma}, \eqref{eq:C_zf_gamma}, and \eqref{eq:C_bd_gamma} can be found at Appendix
\ref{sec:derivation_gamma}. As a result, it is easy to see that the DPC-ZF and DPC-BD offsets are unaffected
by $\gamma_1, \cdots, \gamma_K$:
\begin{theorem}
    The expected loss in Rayleigh fading when each user has a different average SNR is identical to the loss when
    all users have the same average SNR at high SNR. That is,
    \begin{equation}
        \bar{\beta}_\text{DPC-ZF}(M,KN) =\log_2 e \sum_{j=1}^{KN-1} \frac{j}{M-j}\quad \text{(bps/Hz)}, \label{eq:E_beta_zf_gamma}
    \end{equation}
    and
    \begin{equation}
        \bar{\beta}_\text{DPC-BD} (M, K, N) = (\log_2 e)\left(
        \sum_{k=0}^{K-1}\sum_{n=0}^{N-1}\sum_{i=kN+1}^{(K-1)N}\frac{1}{M-n-i}
        \right) \;\;\; \text{\rm (bps/Hz)}, \label{eq:E_beta_bd_gamma}
    \end{equation}
    which are identical with \eqref{eq:E_beta_zf} and \eqref{eq:E_beta_bd}, respectively.
\end{theorem}

\begin{figure}
    \centering
    \includegraphics[width=0.6\textwidth]{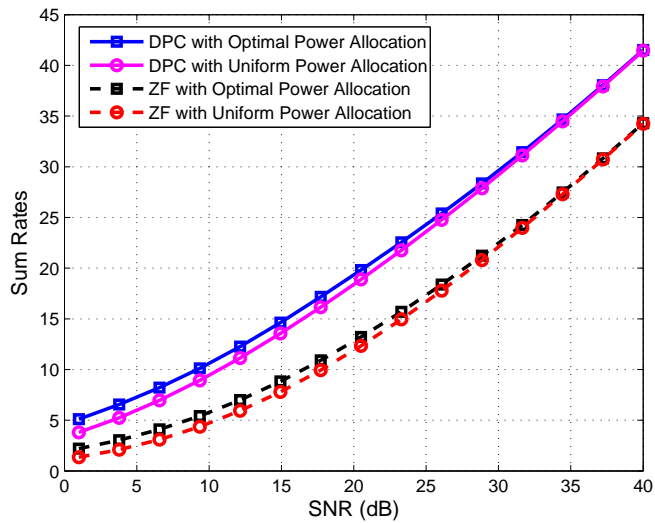}
    \caption{Average sum rates with the optimal power allocation and the uniform power allocation
    when $M=4$, $N=1$, $K=4$, with unequal SNR: $\gamma_1=0.1$, $\gamma_2=0.5$, $\gamma_3=1$, $\gamma_4=2$.}
    \label{fig:unequal_snr_avg_diff}
\end{figure}

Fig.~\ref{fig:unequal_snr_avg_diff} illustrates that the sum rates by optimal power allocation and uniform
power allocation tend to zero as power grows both for DPC and ZF. Unlike the symmetric channel gain case,
more transmit power is required to make the difference sufficiently small.

\section{Weighted Sum Rate Analysis}
\begin{figure}
    \centering
    \includegraphics[width=0.5\textwidth]{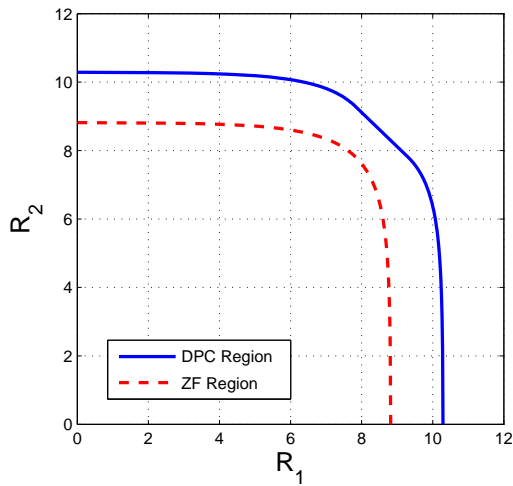}
    \caption{Achievable regions of DPC and ZF for $M=2$, $N=1$, and $K=2$, with ${\bf h}_1 = [1\;\;0.5]$, ${\bf h}_2 = [0.5\;\;1]$ at SNR=30 (dB).}
    \label{fig:cap_region}
\end{figure}

 In this section we generalize the rate offset analysis to weighted
sum rate.  We first consider single antenna receivers ($N=1$), and then discuss the extension to $N>1$ at the
end of this section. Fig.~\ref{fig:cap_region} illustrates the capacity region (DPC) and the ZF achievable
region for a particular 2 user channel at 30 dB. While the sum rate is the point where the negative slope of
the boundary of the rate region is 1, weighted sum rate corresponds to points where the slope of the boundary
are specified by the particular choice of user weights. The sum rate offset quantifies the difference between
the sum rate points of both regions; the weighted sum rate offset is intended to describe the offset for the
other portions of the rate region.

We first show that allocating power in proportion to user weights is
asymptotically optimal for either DPC or ZF, and then use
this result to compute the associated rate offsets.
Then, we show the utility of our simple power allocation
policy via application to queue-based scheduling.

\subsection{Asymptotically Optimal Power Allocation}

Without loss of generality, we assume user weights are in descending order: $\mu_1 \ge \mu_2 \ge \cdots \ge
\mu_K\ge 0$ with $\sum_{k=1}^K \mu_k =1$. The maximum weighted sum rate problem (DPC), which is defined as
the maximum of $\sum_{k=1}^K \mu_k R_k$ over the capacity region, can be written in
terms of the dual MAC as:
\begin{equation} \label{eq:C_diff_mu_modified}
    {\cal C}_{\rm DPC}({\boldsymbol \mu}, {\bf H}, P)=\max_{\sum_{k=1}^K P_k \le P}
    \sum_{k=1}^{K} \mu_k \log_2\left(1+P_k{\bf h}_k({\bf A}^{(k-1)})^{-1} {\bf h}_k^H \right),
\end{equation}
where ${\bf A}^{(k-1)} = {\bf I}+\sum_{j=1}^{k-1} P_j {\bf h}_j^H{\bf h}_j$ for $k\ge 1$ and ${\bf A}^{(0)} =
{\bf I}$. Since $N=1$, each channel is a row vector and is written as ${\bf h}_k$.
Notice that the uplink decoding is done in order of increasing
weight, i.e., user $K$ does not get the benefit of any interference cancellation
while user $1$'s signal benefits from full interference cancellation
and is thus detected in the presence of only noise.

The following lemma shows that if we limit ourselves to
linear power allocation policies of the form $P_k = \alpha_k P$,
then the objective function in (\ref{eq:C_diff_mu_modified})
can be decoupled at high SNR:
\begin{lemma} \label{lem:wt_sum_rate_simple_form}
    If $M\ge K$, then for any $\alpha_k  > 0$, $k=1,\cdots,K$ with $\sum_{k=1}^K \alpha_k =1$,
    \begin{equation}
        \lim_{P\to\infty}\Bigg[\sum_{k=1}^{K} \mu_k \log_2\left(1+\alpha_k P{\bf h}_k({\bf A}^{(k-1)})^{-1} {\bf h}_k^H \right)
        - \sum_{k=1}^{K} \mu_k \log_2\left(1+\alpha_k P\|{\bf f}_k\|^2\right) \Bigg]=0,
    \end{equation}
    where ${\bf f}_k$ is the projection of ${\bf h}_k$ onto the nullspace of $\{{\bf h}_1, \cdots, {\bf
h}_{k-1}\}$.
\end{lemma}
\begin{proof}
    Lemma \ref{lem:decoupling} in Appendix \ref{sec:decoupling_lemma} shows that
${\bf h}_k({\bf A}^{(k-1)})^{-1} {\bf h}_k^H \rightarrow \|{\bf f}_k\|^2$ as $P \rightarrow \infty$.  By the
continuity of $\log(\cdot)$ and the fact that $P \rightarrow \infty$, we get the result.
\end{proof}
Once the weighted sum rate maximization has been decoupled into the problem of maximizing weighted sum rate
over parallel single-user channels, we can use the result of \cite{Lozano_Tulino_Verdu_ISSSTA} to show that
the optimal power allocation is of the form $P_k^* = \mu_k P + O(1)$.

\begin{theorem} \label{thm:wt_sum_rate_pow_policy}
    When $M\ge K$, allocating power according to
    \begin{equation}
        \boxed{P_k = \mu_k P,\qquad k=1,\cdots,K
            \label{eq:wt_sum_rate_pow_policy}}
    \end{equation}
    asymptotically achieves the optimal solution to \eqref{eq:C_diff_mu_modified} at high SNR.
\end{theorem}
\begin{proof}
    By Lemma \ref{lem:wt_sum_rate_simple_form}, the following optimization will yield an
    asymptotically optimal solution
(albeit with a weak restriction on allowable power policies):
    \begin{equation} \label{eq:wt_sum_rate}
        \max_{P_k :\; \sum_{k=1}^{K} P_k\le P}
        \sum_{k=1}^{K} \mu_k\log_2 \left(1+P_k\|{\bf f}_k\|^2 \right).
    \end{equation}
The optimal power policy for a more general version of this problem, in which there are more than $K$
parallel channels and each user can occupy multiple channels,
 is solved in \cite{Lozano_Tulino_Verdu_ISSSTA}.  We need only
consider this simplified version, and it is easily checked (via KKT conditions) that the solution to
\eqref{eq:wt_sum_rate} is:
    \begin{equation} \label{eq-opt_power_weighted}
        P_k^* = \mu_k P + \mu_k \left(\sum_i \frac{1}{\|{\bf f}_i\|^2} \right) -\frac{1}{\|{\bf f}_k\|^2},
        \quad \text{for}\;\; k=1,\cdots,K,
    \end{equation}
    when $P$ is sufficiently large to allow all users to have non-zero power.
Therefore, at high SNR we have
    \begin{equation}
    P_k^* = \mu_k P + O(1) ,\qquad k=1,\cdots,K.
    \end{equation}
    Since the $O(1)$ power term leads to a vanishing rate, we have the result.
\end{proof}
Theorem \ref{thm:wt_sum_rate_pow_policy} generalizes the fact that uniform power allocation achieves the
maximum sum rate asymptotically at high SNR. That is, for the sum rate problem the weights are the same
(i.e., $\mu_1 = \cdots = \mu_K = 1/K$), thus the uniform power policy is asymptotically optimal.

\begin{figure}
        \centering
        \includegraphics[width=0.6\textwidth]{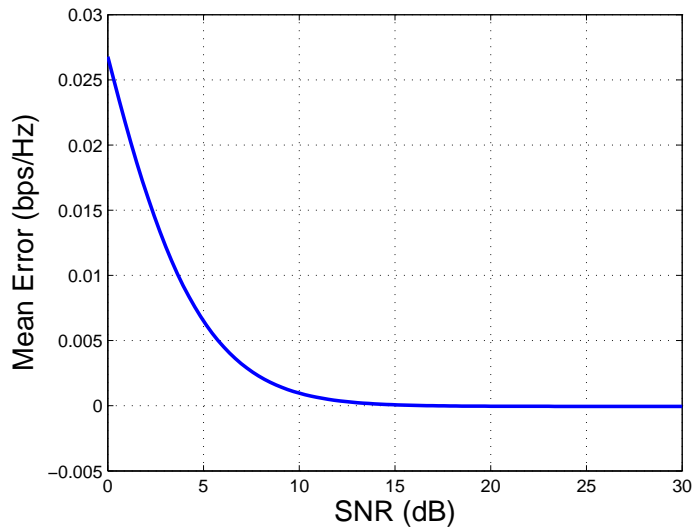}
        \caption{Averaged weighted sum rate difference between the exact solution \eqref{eq:C_diff_mu_modified} and the asymptotic solution \eqref{eq:wt_sum_rate_pow_policy} when $\mu_1=0.6$ and $\mu_2=0.4$ for
        Rayleigh fading channel.}
        \label{fig:diff_mu_high_snr_appr_ex_error_avg}
\end{figure}

In Fig.~\ref{fig:diff_mu_high_snr_appr_ex_error_avg} the difference between the true weighted sum rate
(\ref{eq:C_diff_mu_modified}) and the weighted sum rate achieved using $P_k = \mu_k P$ is plotted as a
function of SNR.  This difference is averaged over iid Rayleigh channel realizations for a $(M=4, K=2, N=1)$
system with $\mu_1=0.6$ and $\mu_2=0.4$.  The approximate power allocation is seen to give a weighted sum
rate that is extremely close to the optimum even at very low SNR values.

Meanwhile, the weighted sum rate by ZF is given by
\begin{equation} \label{eq:wt_sum_rate_zf}
    {\cal C}_\text{ZF}({\boldsymbol \mu},{\bf H},P)=\max_{P_k :\; \sum_{k=1}^{K} P_k\le P}
    \sum_{k=1}^{K} \mu_k\log_2 \left(1+P_k\|{\bf g}_k\|^2 \right),
\end{equation}
where ${\bf g}_k$ is the projection of ${\bf h}_k$ onto the null space of $\{{\bf h}_1, \cdots, {\bf
h}_{k-1}, {\bf h}_{k+1},\cdots, {\bf h}_K\}$.
The result of \cite{Lozano_Tulino_Verdu_ISSSTA} directly applies here,
and therefore the power allocation policy in \eqref{eq:wt_sum_rate_pow_policy} is
also the asymptotic solution to \eqref{eq:wt_sum_rate_zf}.

\subsection{Rate Loss}
Using the asymptotically optimal power allocation policy of (\ref{eq:wt_sum_rate_pow_policy}), the weighted
sum rates of DPC and ZF can be expressed as
\begin{eqnarray}
        {\cal C}_\text{DPC}({\boldsymbol \mu},{\bf H},P) &\cong& \sum_{k=1}^K \mu_k \log\left(1+\mu_k P \|{\bf f}_k\|^2 \right),\\
        {\cal C}_\text{ZF}({\boldsymbol \mu},{\bf H},P) &\cong& \sum_{k=1}^K  \mu_k \log\left(1+\mu_k P \|{\bf g}_k\|^2
        \right).
\end{eqnarray}
Thus, the rate offset per realization is given by
\begin{equation}
    \beta_\text{DPC-ZF}({\boldsymbol \mu},{\bf H})
    \cong \sum_{k=1}^K\mu_k \log\frac{ \|{\bf f}_k\|^2}{ \|{\bf g}_k\|^2}.
\end{equation}
In Rayleigh fading, the distributions of $\|{\bf f}_k\|^2$ and $\|{\bf g}_k\|^2$ are $\chi^2_{2(M-k+1)}$ and
$\chi^2_{2(M-K+1)}$, respectively. Therefore, the expected rate loss is given by
\begin{equation} \label{eq-offset_weighted}
    \bar{\beta}_\text{DPC-ZF}({\boldsymbol \mu}, M, K) \cong
    (\log_2 e)\sum_{k=1}^K\mu_k \left(\sum_{j=M-K+1}^{M-k}\frac{1}{j}\right)\quad.
\end{equation}

It is straightforward to check that the rate offset is minimized at the sum rate, i.e., when
$\mu_1=\cdots=\mu_k=\frac{1}{K}$. If we let $\zeta_k = \sum_{j=M-K+1}^{M-k}\frac{1}{j}$, then
$\zeta_1>\zeta_2>\cdots>\zeta_K$ and  $\bar{\beta}_\text{DPC-ZF} = (\log_2 e) \sum_{k=1}^K \mu_k \zeta_k$.
Since $\{\mu_k\}$ has constraints of $\mu_1\ge \cdots \ge \mu_K$, $\sum_{k=1}^K\mu_1=1$, and $\mu_k\ge 0$
$(1\le k \le K)$, $\bar{\beta}_\text{DPC-ZF}$ achieves minimum at $\mu_1=\cdots=\mu_k=\frac{1}{K}$ for a
given $\{\zeta_k\}$.

\subsection{Application to Queue-based Scheduling}

Queue-based scheduling, introduced by the seminal work of Tassiulas and Ephremides \cite{Tassiulas_AC92}, is
one application in which it is necessary to repeatedly maximize the weighted sum rate for different user
weights.
\begin{figure}
    \centering
    \includegraphics[width=0.60\textwidth]{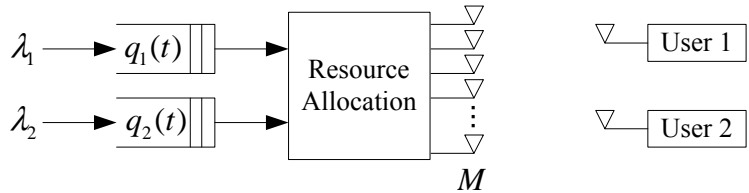}
    \caption{MIMO BC with queues}
    \label{fig:highsnr_w_queue}
\end{figure}
Fig.~\ref{fig:highsnr_w_queue} illustrates a queue-based scheduling system for two users. Data for the users
arrive at rates $\lambda_1$ and $\lambda_2$, which are generally assumed to be unknown. During each time
slot, the transmitter chooses the rate vector that maximizes the weighted sum rate over the instantaneous
rate region with weights equal to the current queue sizes. If the queue lengths are denoted as $q_1(t)$ and
$q_2(t)$, then the transmitter solves the following optimization during each time slot:
\begin{equation}
    \max_{{\bf R}\in {\cal C}({\bf H},P)} q_1(t) R_1 + q_2(t) R_2,
    \label{eq:opt_wgt_queue}
\end{equation}
and such a policy stabilizes any rate vector in the ergodic capacity region.

Although the weighted sum rate maximization problem for DPC stated in equation (\ref{eq:C_diff_mu_modified})
is convex, it still requires considerable complexity and could be difficult to perform on a slot-by-slot
basis.  An alternative is to use the approximate power allocation policy from
(\ref{eq:wt_sum_rate_pow_policy}) during each time slot:
\begin{eqnarray}
    P_k &=& \frac{q_k(t)}{q_1(t)+q_2(t)} P, \label{eq:queue_pow_alloc_K2_P1}
\end{eqnarray}
and where the ordering of the queues determines the dual MAC decoding order (larger queue decoded last).

Although we do not yet have any analytical results on the performance of the asymptotically optimal power
policy, numerical results indicate that such a policy performs nearly as well as actually maximizing weighted
sum rate. Ongoing work is investigating whether the approximate strategy is stabilizing for this system.

\begin{figure}
    \centering
    \includegraphics[width=0.6\textwidth]{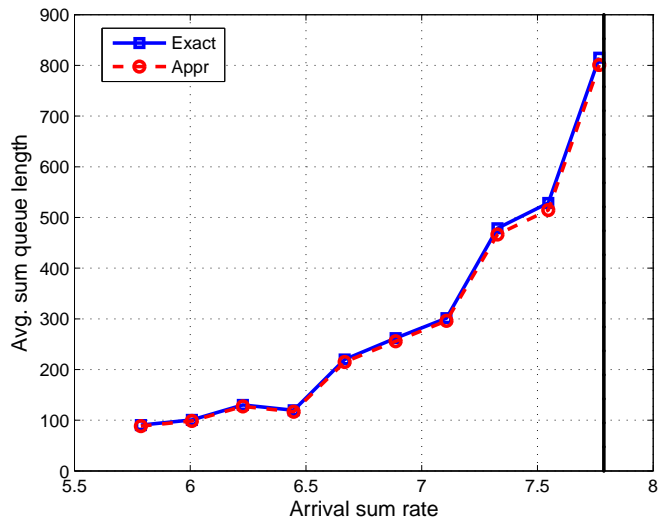}
    \caption{Average queue length for symmetric arrival.}
    \label{fig:queue_length_vs_lambda}
\end{figure}

In Fig.~\ref{fig:queue_length_vs_lambda} average queue length is plotted versus the sum arrival rate for an
$M=4$, $K=2$ channel at 10 dB, for both the exact weighted sum rate maximization as well as the
approximation.  Both schemes are seen to perform nearly identical, and the approximate algorithm appears to
stabilize the system in this scenario, although this is only empirical evidence.

\subsection{Extension to $N>1$}

Similar to \eqref{eq:C_diff_mu_modified}, the weighted sum rate by DPC can be as:
\begin{equation} \label{eq:C_diff_mu_modified_N}
    {\cal C}_{\rm DPC}({\boldsymbol \mu}, {\bf H}, P)=\max_{\sum_{k=1}^K \text{tr}({\bf Q}_k) \le P}
    \sum_{k=1}^{K} \mu_k \log_2 \frac{|{\bf A}^{(k)}|}{|{\bf A}^{(k-1)}|},
\end{equation}
where ${\bf A}^{(k)}= {\bf I} + \sum_{j=1}^k {\bf H}_j^H {\bf Q}_j {\bf H}_j$ for $k\ge 1$ and ${\bf
A}^{(0)}= {\bf I}$. From the construction of ${\bf A}^{(k)}$,
\begin{equation*}
    \frac{|{\bf A}^{(k)}|}{|{\bf A}^{(k-1)}|} = \left|{\bf I}+{\bf Q}_k{\bf H}_k ({\bf A}^{(k-1)})^{-1} {\bf H}_k^H\right|.
\end{equation*}
Hence, \eqref{eq:C_diff_mu_modified_N} can be written as
\begin{equation} \label{eq:C_diff_mu_modified1_N}
    {\cal C}_{\rm BC}({\boldsymbol \mu}, {\bf H}, P)=\max_{\sum_{k=1}^K \text{tr}({\bf Q}_k) \le P}
    \sum_{k=1}^K \mu_k \log_2 \left|{\bf I}+{\bf Q}_k{\bf H}_k ({\bf A}^{(k-1)})^{-1} {\bf H}_k^H\right|.
\end{equation}
With the decoupling lemma (see Appendix \ref{sec:decoupling_lemma}), the above optimization
\eqref{eq:C_diff_mu_modified1_N} can be solved asymptotically like the case of $N=1$:
\begin{theorem} \label{thm:C_diff_mu_N}
    At high SNR, the optimization in \eqref{eq:C_diff_mu_modified_N} is asymptotically achieved when
    \begin{equation} \label{eq:opt_sol_C_diff_mu_N}
        \boxed{{\bf Q}_k = \frac{\mu_k P}{N} {\bf I}, \qquad k=1,\cdots, K.}
    \end{equation}
\end{theorem}
\begin{proof}
    See Appendix \ref{sec:pf_thm:C_diff_mu_N}.
\end{proof}

Similarly, the weighted sum rate of BD is given by
\begin{equation} \label{eq:wt_sum_rate_bd}
    {\cal C}_\text{BD}({\boldsymbol \mu},{\bf H},P)=\max_{{\bf Q}_k \; : \;\sum_{k=1}^{K} \text{tr}({\bf Q}_k)\le P}
    \sum_{k=1}^{K} \mu_k\log_2 \left|{\bf I}+{\bf Q}_k {\bf G}_k {\bf G}_k^H\right|,
\end{equation}
where ${\bf G}_k$ is the projection of ${\bf H}_k$ onto the null space of $\{{\bf H}_1, \cdots, {\bf
H}_{k-1}, {\bf H}_{k+1},\cdots, {\bf H}_K\}$. Likewise, the optimization \eqref{eq:wt_sum_rate_zf} is the
same as the optimization \eqref{eq:wt_sum_N_sep1} and \eqref{eq:wt_sum_N_sep2} except that ${\bf F}_k$ is
replaced by ${\bf G}_k$ which does not contribute to the asymptotic solution. Thus, the power allocation
policy in \eqref{eq:opt_sol_C_diff_mu_N} is also the asymptotic solution to \eqref{eq:wt_sum_rate_bd}.

\subsection{More Users Than Antennas}

Although it is asymptotically optimal to allocate power in proportion to user weights when $M \geq KN$, this
is not the case when $M < KN$. Indeed, such a strategy can easily be checked to be sub-optimal even for a
single antenna broadcast channel with more than one user, as considered in
\cite{Li_IT01}\cite{Tse_unpublished}. Allocating power directly proportional to user weights or allocating
all power to only the user with the largest weight yields, for many single antenna broadcast channels, a
weighted sum rate that is a bounded distance away from the true optimal weighted sum rate.

Although neither of these strategies is asymptotically optimal, numerical
results do show that these approximations achieve rates that are extremely
close to optimum. In general, there are two different reasonable
power approximations.   The first is to simply choose $P_k = \mu_k P$.
However, when $K > M$, this results in sending many more data streams than
there are spatial dimensions, which is not particularly intuitive.
An alternative strategy is to allocate power to the users with the
$M$ largest weights, but again in proportion to their weights.

\begin{figure}
    \centering
    \includegraphics[width=0.6\textwidth]{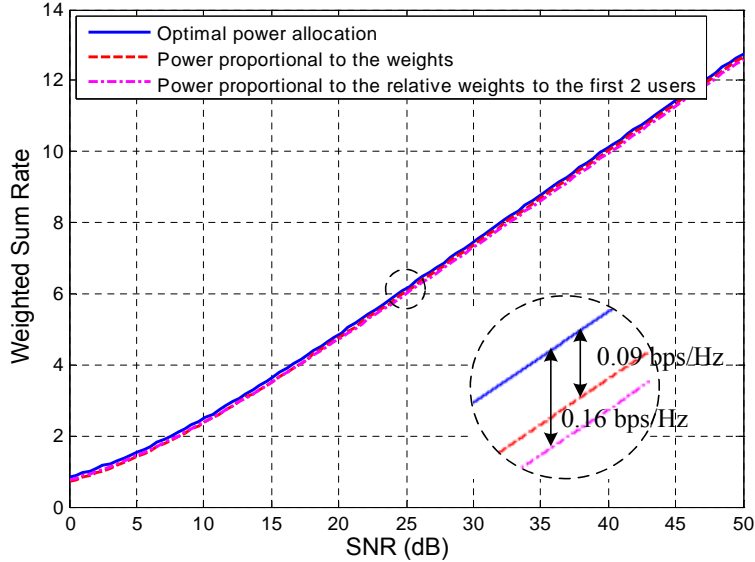}
    \caption{Ergodic weighted sum rates by DPC and by approximations when $M=2$, $N=1$, $K=3$, with $\mu_1=0.5$, $\mu_2=0.3$, and $\mu_3=0.2$.}
    \label{fig:ergordic_wt_sum}
\end{figure}

Fig.~\ref{fig:ergordic_wt_sum} illustrates the ergodic weighted sum rates vs SNR for a $K=3, M=2, N=1$ system
in which $\mu_1=0.5$, $\mu_2=0.3$, and $\mu_3=0.2$, averaged over Rayleigh fading. The true weighted sum rate
is compared to the first strategy, where $P_k = \mu_k P$, and to the second strategy, where only users 1 and
2 are allocated power according to: $P_1=\frac{\mu_1}{\mu_1+\mu_2} P$, $P_2=\frac{\mu_2}{\mu_1+\mu_2} P$, and
$P_3=0$. Both approximations are a non-zero distance away from the optimum, but the rate loss is seen to be
extremely small.

\section{Conclusion}

We have investigated the difference between the throughputs achieved by dirty paper coding (DPC) relative to
those achieved with linear precoding strategies by utilizing the affine approximation to high SNR and
computing the exact throughput/power offsets at asymptotically high SNR for MIMO broadcast channels in which
the number of transmit antennas is no smaller than the total number of receive antennas. Simple expressions
in terms of the number of transmit and receive antennas are provided for the average rate/power offset in a
spatially white Rayleigh fading environment. When the aggregate number of receive antennas is equal or
slightly less than than the number of transmit antennas, linear precoding incurs a rather significant penalty
relative to DPC, but this penalty is much smaller when the number of transmit antennas is large relative to
the number of receive antennas.

Furthermore, we generalized our analysis to weighted sum rate and quantified the asymptotic rate/power
offsets for this scenario as well.  One of the most interesting aspects of this extension is the finding that
allocating power directly proportional to user weights is asymptotically optimal for DPC at high SNR. 
This result is an extension at a similar result for parallel channels found in
\cite{Lozano_Tulino_Verdu_ISSSTA}, and this simple yet asymptotically optimal power policy may prove to be
useful in other setting such as opportunistic scheduling.

\section*{Acknowledgment}
The authors would like to thank Angel Lozano for his comments, and in particular for suggesting the unequal
average SNR model.

\appendices

\section{Proof of Theorem \ref{thm:exp_beta_zf}}
\label{sec:pf_thm:exp_beta_zf}

Starting from (\ref{eq-dpc_zf_proof1}) and utilizing
$\psi(m) = \psi(1) + \sum_{l=1}^{m-1} \frac{1}{l}$ we have:
\begin{eqnarray*}
    \E\left[\log_e \beta_\text{DPC-ZF} \right]
    &=& \left( \sum_{l=0}^{KN-1}\psi(M-l) \right) - KN \psi(M-KN+1) \\
    &=& \sum_{l=0}^{KN-1} \left(\sum_{j=1}^{M-l-1}\frac{1}{j}-\sum_{j=1}^{M-KN}\frac{1}{j} \right)\\
    &=& \sum_{l=0}^{KN-1} \left(\sum_{j=M-KN+1}^{M-1}\frac{1}{j} \right)\\
    &=& \sum_{j=1}^{KN-1} \frac{KN-j}{M-KN+j}\\
    &=& \sum_{j=1}^{KN-1} \frac{j}{M-j}.
\end{eqnarray*}

\section{Derivation of \eqref{eq:beta_zf_M_K_equal}}
\label{sec:pf_eq:beta_zf_M_K_equal}
When $M=KN$,
\begin{equation}
    \E_{\bf H}[\log_e \beta_\text{DPC-ZF}({\bf H})]
    = \sum_{j=1}^{KN-1}\frac{j}{M-j}=\sum_{j=1}^{M-1}\sum_{i=1}^{M-j}\frac{1}{i}.
\end{equation}
If we let $S_M$ denote the expected rate loss with $M$ antennas, we have:
\begin{equation}
    S_{M+1}-S_M = \sum_{i=1}^M\frac{1}{i} \le 1+\log M , \quad \text{for}\;\; M\ge 1 ,
\end{equation}
since $\log_e M = \int_1^M \frac{1}{x} dx \ge \sum_{i=2}^M \frac{1}{i}$ because $\frac{1}{x}$ is a decreasing
function. If we let $f(M)\triangleq M\log_e M$, $f'(M)=1+\log_e M$, which is an increasing function of $M$,
and thus $f(M+1)\ge f(M)+1+\log_e M$. Since $S_{M+1}-S_M\le 1+\log M$ and $f(1)=S_1=0$, $S_M\le M\log M$ for
all $M\ge 1$.

Now we show that $S_M$ converges to $M\log M$. We do this by showing that $S_M\ge \theta M \log_e M$ for any
$0<\theta<1$ for all $M$ larger than some $M_0$. First notice that $\log_e M \le \sum_{i=1}^{M-1} \frac{1}{i}
\le \sum_{i=1}^M \frac{1}{i}$ by the definition of the $\log(\cdot)$ function. Thus,
\begin{equation}
    S_{M+1}-S_M = \sum_{i=1}^M \frac{1}{i} \ge \log_e M.
\end{equation}
Let $g(M)\triangleq \theta M \log M$ for some $0<\theta<1$. Then $g'(M)=\theta+\theta \log_e M$, which is an
increasing function of $M$. Thus $g(M+1)\le g(M)+g'(M+1)=g(M)+\theta+\theta \log_e(M+1)$. Therefore we have
\begin{equation}
    g(M+1)-S_{M+1} \le \left(f(M)-S_M\right)
    +\theta+\theta\log(M+1)-\log M.
\end{equation}
Notice that the term $\theta+\theta\log(M+1)-\log M$ is a monotonically decreasing function that goes to
$-\infty$. Thus, any positive gap between $g(M)$ and $S_M$ must close and go to $-\infty$, i.e., $S_M\ge
g(M)$ for sufficiently large $M$. As a consequence of this, $\lim_{M\to\infty}\frac{S_M}{\theta M \log_e
M}\ge 1$, or $\lim_{M\to\infty}\frac{S_M}{M\log_e M}\ge \theta$ for any $\theta<1$. Since $\frac{S_M}{M\log_e
M}$ is bounded above by 1, it must converge; i.e.,
\begin{equation}
    \lim_{M\to\infty}\frac{S_M}{M\log_e M}=1
\end{equation}
as desired.

\section{Proof of Theorem \ref{thm:asymp_zf_penalty}}
\label{sec:pf_thm_asymp_zf_penalty}
    From Theorem \ref{thm:exp_beta_zf}, if $M=\alpha KN$, the expected power
    offset, which is now a function of $\alpha$ and $KN$, can be expressed as:
    \begin{eqnarray*}
        \bar{\Delta}_\text{DPC-ZF}(\alpha,KN)&=& \frac{3\log_2 e}{KN}\sum_{j=1}^{KN-1}\frac{j}{M-j}, \qquad M=\alpha KN \\
        &=&3\log_2 e \sum_{j=1}^{KN-1}\frac{\frac{j}{KN}}{\alpha-\frac{j}{KN}}\frac{1}{KN}
    \end{eqnarray*}
    Let us define a function $f(x)$ as
    \begin{equation*}
        f(x) = \frac{x}{\alpha-x},\qquad x\in [0,1],\quad\alpha > 1.
    \end{equation*}
    Then $\bar{\Delta}_\text{DPC-ZF}$ can be expressed as
    \begin{equation*}
        \bar{\Delta}_\text{DPC-ZF}(\alpha,KN)
        =3\log_2 e\sum_{j=1}^{KN-1}f\left(\frac{j}{KN}\right)\frac{1}{KN},
    \end{equation*}
    which is a Riemann sum; i.e., as $KN\to\infty$,
    \begin{equation*}
        \lim_{KN\to\infty} \bar{\Delta}_\text{DPC-ZF}(\alpha,KN)
        =3\log_2 e\int_{0}^{1}f(x)dx=3\log_2 e\int_{0}^{1}\frac{x}{\alpha-x}dx.
    \end{equation*}
    Thus,
    \begin{eqnarray}
        \bar{\Delta}_\text{DPC-ZF}(\alpha)=
        \lim_{KN\to\infty} \bar{\Delta}_\text{DPC-ZF}(\alpha,KN)
        &=& 3\log_2 e\left(-1-\alpha\log_e\frac{\alpha-1}{\alpha} \right) \nonumber\\
        &=& -3 \left(\log_2 e+\alpha \log_2 \left(1-\frac{1}{\alpha} \right) \right) \nonumber.
    \end{eqnarray}

\section{Proof of Theorem \ref{thm:exp_beta_bd}}
\label{sec:pf_thm:exp_beta_bd}

From \eqref{eq:sum_rate_dpc_appr} and \eqref{eq:sum_rate_bd_appr},
$\bar{\beta}_\text{DPC-BD}$ is given by:
    \begin{equation*}
        \bar{\beta}_\text{DPC-BD}
        = \E[\log_2 \left|{\bf H}^H {\bf H} \right|]-K \E\left[\log_2 \left| {\bf G}_k^H {\bf G}_k\right| \right]
    \end{equation*}
where ${\bf G}_k^H {\bf G}_k$ is Wishart wth $M-(K-1)N$ degrees of freedom. Applying Lemma \ref{lem-wishart}
and expanding the digamma function we have:
    \begin{eqnarray*}
        \bar{\beta}_\text{DPC-BD} &=&\log e \left[ \sum_{l=0}^{KN-1}\psi(M-l)-K \sum_{n=0}^{N-1} \psi(M-(K-1)N-n)  \right]\\
        &=& \log e \sum_{k=0}^{K-1}\sum_{n=0}^{N-1}\left[\psi(M-n-kN)- \psi(M-n-(K-1)N)  \right]\\
        &=& \log e \sum_{k=0}^{K-1}\sum_{n=0}^{N-1}\left[\sum_{j=1}^{M-n-kN-1}\frac{1}{j}- \sum_{j=1}^{M-n-(K-1)N-1}\frac{1}{j}  \right]\\
        &=& \log e \sum_{k=0}^{K-1}\sum_{n=0}^{N-1}\sum_{i=kN+1}^{(K-1)N}\frac{1}{M-n-i}.
    \end{eqnarray*}

\section{Proof of Theorem \ref{thm:diff_beta}}
\label{sec:pf_thm:diff_beta}
    From \eqref{eq:E_beta_zf} and \eqref{eq:E_beta_bd} it is known
    that the $\bar{\beta}_\text{DPC-BD}$ and $\bar{\beta}_\text{DPC-ZF}$ are given by
    \begin{eqnarray*}
        \bar{\beta}_\text{DPC-BD} &=&\log e \left[ \sum_{l=0}^{KN-1}\psi(M-l)-K \sum_{n=0}^{N-1} \psi(M-(K-1)N-n) \right],\\
        \bar{\beta}_\text{DPC-ZF} &=&\log e \left[ \sum_{l=0}^{KN-1}\psi(M-l)-KN \psi(M-KN+1)\right].
    \end{eqnarray*}
    From the assumption, $M=\alpha KN$ ($\alpha>1$) with $N>1$,
    \begin{eqnarray*}
        \frac{1}{\log e}\left(\E[\beta_\text{DPC-ZF}]-\E[\beta_\text{DPC-BD}]\right)
        &=& K\left(\sum_{n=0}^{N-1} \psi(M-(K-1)N-n)\right)-KN \psi(M-KN+1)  \\
        &=& K\sum_{n=2}^{N} \left[\psi(M-KN+n)- \psi(M-KN+1)\right]  \\
        &=& K\sum_{n=2}^{N} \sum_{j=M-KN+1}^{M-KN+n-1}\frac{1}{j} \\
        &=& K\sum_{i=1}^{N-1} \frac{N-i}{M-KN+i}\\
        &=& \sum_{i=1}^{N-1} \frac{K(N-i)}{(\alpha-1)KN+i}
    \end{eqnarray*}

\section{Derivation of \eqref{eq:C_dpc_gamma}, \eqref{eq:C_zf_gamma}, and \eqref{eq:C_bd_gamma}}
\label{sec:derivation_gamma}
From \eqref{eq:sum_cap} with the uniform power allocation, we have
\begin{equation*}
    {\cal C}_\text{DPC}({\bf H},P) \cong \log_2 \left|{\bf I}+\frac{P}{KN}\tilde{\bf H}^H{\boldsymbol \Gamma}\tilde{\bf H}
    \right|,
\end{equation*}
where ${\boldsymbol \Gamma} = \text{diag}(\gamma_1,\cdots,\gamma_K )\otimes {\bf I}_{N\times N}$. By $|{\bf
I}+{\bf A}{\bf B}|=|{\bf I}+{\bf B}{\bf A}|$,
\begin{eqnarray*}
    {\cal C}_\text{DPC}({\bf H},P) &\cong& \log_2 \left|{\bf I}+\frac{P}{KN}{\boldsymbol \Gamma}\tilde{\bf H}\tilde{\bf H}^H
    \right|\\
    &=& KN\log_2 P + \log_2 \left|\frac{1}{P}{\bf I}+\frac{1}{KN}{\boldsymbol \Gamma}\tilde{\bf H}\tilde{\bf H}^H \right|\\
    &\cong& KN\log_2 P + \log_2 \left|\frac{1}{KN}{\boldsymbol \Gamma}\tilde{\bf H}\tilde{\bf H}^H \right|\\
    &=& KN\log_2 P -KN\log_2 KN +\log_2 \left|{\boldsymbol \Gamma} \right| + \log_2 \left|\tilde{\bf H}\tilde{\bf H}^H \right|\\
    &\cong& {\cal C}_\text{DPC}(\tilde{\bf H},P) + N \sum_{k=1}^K \log_2 \gamma_k
\end{eqnarray*}

Since the zero-forcing vector ${\bf v}_{k,n}$ for ${\bf h}_{k,n}$ is identical to the zero-forcing vector
$\tilde{\bf v}_{k,n}$ for $\tilde{\bf h}_{k,n}$, the effective channel gain is given by
\begin{equation}
    g_{k,n}={\bf h}_{k,n}{\bf v}_{k,n} = \sqrt{\gamma_k}\tilde{g}_{k,n},
\end{equation}
where $\tilde{g}_{k,n} = \tilde{\bf h}_{k,n}\tilde{\bf v}_{k,n} $. Thus the ZF sum rate \eqref{eq:c_sum_zf}
can be modified as
\begin{eqnarray*}
    {\cal C}_\text{ZF}({\bf H},P)
    &\cong& \sum_{k=1}^K \sum_{n=1}^N \log_2 \left(1+\frac{P}{KN} \gamma_k |\tilde{g}_{k,n}|^2\right)\\
    &=& KN\log_2 P+ \sum_{k=1}^K \sum_{n=1}^N \log_2 \left(\frac{1}{P}+\frac{1}{KN} \gamma_k |\tilde{g}_{k,n}|^2\right)\\
    &\cong& KN\log_2 P+ \sum_{k=1}^K \sum_{n=1}^N \log_2 \left(\frac{1}{KN} \gamma_k |\tilde{g}_{k,n}|^2\right)\\
    &\cong& {\cal C}_\text{ZF}(\tilde{\bf H},P) + N\sum_{k=1}^K \log_2 \gamma_k
\end{eqnarray*}

Likewise, for BD, $\tilde{\bf V}_k = {\bf V}_k$ leads to
\begin{equation}
    {\bf G}_k = {\bf H}_k {\bf V}_k = \sqrt{\gamma_k} \tilde{\bf G}_k,
\end{equation}
where $\tilde{\bf G}_k =\tilde{\bf H}_k \tilde{\bf V}_k$. Thus, the BD sum rate in \eqref{eq:sum_bd} is
modified to
\begin{eqnarray*}
    {\cal C}_\text{BD}({\bf H},P)
    &\cong& \sum_{k=1}^K \log_2 \left|{\bf I}+\frac{P}{KN} \gamma_k \tilde{\bf G}_k^H \tilde{\bf G}_k\right|\\
    &=& KN\log_2 P+ \sum_{k=1}^K \log_2 \left|\frac{1}{P}{\bf I}+\frac{1}{KN} \gamma_k \tilde{\bf G}_k^H \tilde{\bf G}_k\right|\\
    &\cong& KN\log_2 P+ \sum_{k=1}^K \log_2 \left|\frac{1}{KN} \gamma_k \tilde{\bf G}_k^H \tilde{\bf G}_k\right|\\
    &\cong& {\cal C}_\text{BD}(\tilde{\bf H},P) + N\sum_{k=1}^K\log_2 \gamma_k
\end{eqnarray*}

\section{Decoupling Lemma}
\label{sec:decoupling_lemma}
\begin{lemma} \label{lem:decoupling}
    Let $\{{\bf H}_j\}_{j=1}^{K} (\in \mathbb{C}^{N\times M})$ be a set of $K$-user MIMO broadcast channel matrices with $M\ge KN$.
    If\; ${\bf F}_k$ ($k=1,\cdots, K$) is the projection of\; ${\bf H}_k$ onto the nullspace of $\{{\bf H}_j\}_{j=1}^{k-1}$ (i.e., ${\bf F}_k={\bf H}_k {\bf P}^\perp$ where
    ${\bf P}^\perp$ denotes the nullspace of $\{{\bf H}_j\}_{j=1}^{k-1}$), then
    \begin{equation}
        \lim_{P\to\infty}\left[{\bf H}_k ({\bf A}^{(k-1)})^{-1} {\bf H}_k^H - {\bf F}_k {\bf F}_k^H  \right]=0,\qquad k=1,\cdots, K,
    \end{equation}
    where ${\bf A}^{(k)}= {\bf I} + \sum_{j=1}^k {\bf H}_j^H {\bf Q}_j {\bf H}_j$ for $k\ge 1$ and ${\bf A}^{(0)}= {\bf I}$.
\end{lemma}
\begin{proof}
    If we let the eigenvector matrix and eigenvalues of $\sum_{j=1}^{k-1} {\bf H}_j^H {\bf Q}_j {\bf H}_j$ be
    ${\bf U}$ and $\lambda_1, \cdots, \lambda_{k-1}$ with $\lambda_j > 0$, then
    \begin{equation*}
        ({\bf A}^{(k-1)})^{-1/2} = {\bf U}{\boldsymbol \Lambda}{\bf U}^H,
    \end{equation*}
    where
    \begin{equation*}
        {\boldsymbol \Lambda} =\text{diag}\left(\frac{1}{\sqrt{1+\lambda_1}}, \cdots, \frac{1}{\sqrt{1+\lambda_{k-1}}},1,\cdots,1 \right).
    \end{equation*}
    As $P$ goes to infinity, $\lambda$'s tend to infinity. Thus, the first $k-1$ eigenvalues of ${\boldsymbol
    \Lambda}$ converge to 0. The eigenvectors corresponding to the unit eigenvalues span the nullspace $\{{\bf
    H}_j\}_{j=1}^{k-1}$; i.e.,
    \begin{equation*}
        \lim_{P\to\infty} \left[{\bf H}_k({\bf A}^{(k-1)})^{-1/2} - {\bf F}_k  \right] =0.
    \end{equation*}
    This completes the proof.
\end{proof}

\section{Proof of Theorem \ref{thm:C_diff_mu_N}}
\label{sec:pf_thm:C_diff_mu_N}
With the decoupling lemma in Appendix \ref{sec:decoupling_lemma}, the optimization
\eqref{eq:C_diff_mu_modified1_N} can be decomposed into the two optimizations at high SNR:
\begin{equation}
    {\cal C}_{\rm BC}({\boldsymbol \mu}, {\bf H}, P)\cong \max_{\sum_{k=1}^K P_k \le P}
    \sum_{k=1}^K \mu_k \xi_k(P_k),
    \label{eq:wt_sum_N_sep1}
\end{equation}
where
\begin{equation}
    \xi_k(P_k) = \max_{\text{tr}({\bf Q}_k)=P_k} \log_2 \left|{\bf I}+{\bf Q}_k{\bf F}_k{\bf F}_k^H \right|.
    \label{eq:wt_sum_N_sep2}
\end{equation}
At high SNR, Eq.~\eqref{eq:wt_sum_N_sep2} can be asymptotically expressed as an affine approximation
\cite{Shamai_IT01}:
\begin{equation}
    \xi_k(P_k) = {\cal S}_{\infty,k} (\log_2 P_k - {\cal L}_{\infty,k}) + o(1),
\end{equation}
where ${\cal S}_{\infty,k}$ and ${\cal L}_{\infty,k}$ are determined by the multiplexing gain and power
offset. Hence, the optimization \eqref{eq:wt_sum_N_sep1} is asymptotically equivalent to solve the following:
\begin{equation*}
    \max_{\sum_{k=1}^K P_k \le P}\sum_{k=1}^K \mu_k \log_2 P_k.
\end{equation*}
This leads the optimal $P_k=\mu_k P$. Furthermore, by Theorem 3 of \cite{Caire_IT03}, the optimal power
allocation is asymptotically achieved by
\begin{equation*}
    {\bf Q}_k =\frac{\mu_k P}{N}{\bf I},\qquad k=1,\cdots, K.
\end{equation*}

\bibliographystyle{ieeetran} \bibliography{high_snr}

\end{document}